\documentclass[preprint]{revtex4}
\usepackage{graphicx}

\begin{document}
\title{Superfluid-Mott Insulator Transition of Spin-2 Cold Bosons in an Optical Lattice in a Magnetic Field}

\author{Shuo Jin$^1$\footnote{Electronic address: jinshuo@eyou.com},  Jing-Min Hou$^1$
\footnote{Electronic address: jmhou@eyou.com}, Bing-Hao Xie$^2$,
Li-Jun Tian$^1$, and Mo-Lin Ge$^1$}

 \affiliation{$^1$Theoretical Physics
Division, Nankai Institute of Mathematics, Nankai University,
Tianjin, 300071, China, and\\ Liuhui Center for Applied
Mathematics,
Tianjin, 300071, China\\
 $^2$Laboratory for Computational Physics, Institute of Applied Physics and
Computational Mathematics, Beijing, 100088, China}

\begin{abstract}
The superfluid-Mott insulator transition of spin-2 boson atoms
with repulsive interaction in an optical lattice in a magnetic
field is presented. By using the mean field theory, Mott ground
states and phase diagrams of superfluid-Mott insulator transition
at zero temperature are revealed. Applied magnetic field leads to
some phase boundaries splitting. For all the initial Mott ground
states containing multiple spin components, different spin
components take on different phase boundaries. It is found that in
this system the phase boundaries with different magnetization can
be moved in different ways by only changing the intensity of the
applied magnetic
field.\\
PACS number(s): 03.75.Kk, 03.75.Lm, 03.75.Mn, 32.80.Pj
\end{abstract}

\maketitle

\vspace{10cm}
\section{Introduction}
Recent remarkable experiments
\cite{anderson,orzel,greiner,greiner1} on the superfluid (SF) to
Mott insulator (MI) transition in a system of ultracold atoms in
an optical lattice open intriguing prospects for studying
many-body phenomena, associated with strongly correlated systems
in a highly controllable environment. The optical lattices
\cite{Grynberg,jessen}---arrays of microscopic potentials induced
by the ac stark effect of interfering laser beams---provide ideal
conditions for the study of the laser cooling and the quantum
phase transitions of the confined cold atoms. The dynamics of the
confined atoms in optical lattices is adequately described by the
Bose-Hubbard model \cite{jaksch,fisher}, which predicts SF-MI
transition at low temperature with increasing the ratio of the
on-site interaction to the hopping matrix element. Besides many
experimental efforts made to realize SF-MI transition, a large
number of theoretical studies have appeared
\cite{fisher,jaksch,oosten,chen}. In reference \cite{oosten} an
appropriate mean-field approximation was developed for the
Hamiltonian of spinless or polarized bosons in an optical lattice,
and in order to describe the zero-temperature phase transition
from the superfluid to the Mott-insulating phase, the phase
diagrams were calculated.

Since optical traps \cite{stamper-kurn,stenger,barrett} liberate
the spin degrees of freedom and make possible condensation of
spinor bosons, extensive interests have been stimulated in the
study of multi-component spinor BEC. The quantum phase transition
in spinor BEC, as well as a variety of other novel phenomena
\cite{ho,ohmi,ciobanu,koashi,ueda} were well studied.
Subsequently, inspired by these works, the theoretical researches
about the SF-MI transition of the spinor bosons trapped in an
optical lattice arise. Most recently, Demler and Zhou
\cite{demler} have studied spin-1 Bose atoms in an optical lattice
and obtained several unique properties. Tsuchiya \textit{et al}.
\cite{tsuchiya}, Hou and Ge \cite{hou} have investigated the
spin-1 and spin-2 bosons in an optical lattice with the mean-field
approximation method and obtained the phase diagrams showing a
transition from Mott insulator to superfluid, respectively.

On the other hand, the response to external magnetic field of BEC
is also a topic with interests
\cite{bulgakov,duine,zhang,genkin,koashi,ueda}. The
experimentalists have concentrated on investigating the systems in
an applied magnetic field  because the phase transition can be
tuned by adjusting the magnetic field rather than changing the
samples measured \cite{yazadni,van,hebard,pannetier}. Thus the
theoretical study in this aspect is necessary. So far, Ueda and
Koashi \cite{koashi,ueda} have discussed the magnetic response of
spin-1 and spin-2 BEC in a mesoscopic regime; as to the system in
an optical lattice, Svicainsky and Chui \cite{chui} have studied
the spin-1 bosons in a magnetic field and shown some effects
induced by the magnetic field. However, the spin-2 case in a
magnetic field has not been discussed yet. What about the ground
states, quantum phase transition and the influence of the magnetic
field on this system? This is our mission in the present letter.
First, ignoring the hopping term of the Hamiltonian, we get the
site-independent Hamiltonian, its energy eigenvalues and the Mott
ground states for different cases. Then, applying mean-field
approximation and regarding the hopping term as a perturbation, we
perform the calculations in second-order and draw the phase
diagrams. The response of the phase diagrams to the applied
magnetic field is qualitatively analyzed subsequently. Finally, we
give some remarks and the conclusion.

\section{The Model}
We consider a dilute gas of boson atoms with hyperfine spin $F=2$,
such as $^{23}{\rm Na}$,$ ^{87}{\rm Rb}$ or $^{85}{\rm Rb}$
subject to an external magnetic field in an optical lattice. Based
on Ref. \cite{hou}, the Hamiltonian of spin-2 bosons with
repulsive interaction in an optical lattice including a magnetic
field term can be written in the second-quantized form:
\begin{eqnarray}
\label{h} H=H_A+H_B,
\end{eqnarray}
\begin{eqnarray}
H_A=\int d{\bf r}[\frac{\hbar}{2M}{\bf
\nabla}\Psi_\alpha^\dag\cdot{\bf \nabla}\Psi_\alpha+V({\bf r})
\Psi_\alpha^\dag\Psi_\alpha
-\bar\mu\Psi_\alpha^\dag\Psi_\alpha+\nonumber
\frac{\bar{c}_0}{2}\Psi_\alpha^\dag\Psi_\beta^\dag\Psi_\beta\Psi_\alpha
\\+\frac{\bar{c}_1}{2}\sum_i(\Psi_\alpha^\dag(F_i)_{\alpha\beta}\Psi_\beta)^2
+\bar{c}_2\Psi_\alpha^\dag\Psi_{\alpha^\prime}^\dag\langle
2\alpha;2\alpha^\prime|00\rangle\langle00|2\beta;2\beta^\prime\rangle\Psi_\beta\Psi_{\beta^\prime}],
\label{a}
\end{eqnarray}
\begin{eqnarray}
\label{b} H_B=-\mu_Bg\int d{\bf
r}\Psi_\alpha^\dag({\bf{B}}\cdot{\bf{F}})_{\alpha,\beta}\Psi_\beta,
\end{eqnarray}
 where $V({\bf
r})=V_0(\sin^2kx+\sin^2ky+\sin^2kz)$ with $k$ the wave vector of
the laser light and $V_0$ a tunable amplitude characterizing an
optical lattice, $B$ is a uniform magnetic field independent of
time, $\mu _B$ is the Bohr magneton and $g$ is the Lande factor of
an atom,
 $M$ is the atomic mass, $\Psi_{+2},...,\Psi_{-2}$ are the
five-component field operators corresponding to the sublevels
$m_F=+2,...,-2$ of the hyperfine state $F=2$, $\bar\mu$ is the
chemical potential, $\bar{c_0}=4\pi \hbar^2(3a_4+4a_2)/7m$,
$\bar{c_1}=4\pi \hbar^2(a_4-a_2)/7m$ and $\bar{c_2}=4\pi
\hbar^2(3a_4-10a_2+7a_0)/7m$ are parameters related to s-wave
scattering lengths $a_0$, $a_2$ and $a_4$ of the two colliding
bosons with total angular momenta $0, 2$ and $4$,
$<2\alpha;2\alpha^\prime|00>$ and $<00|2\beta;2\beta^\prime>$ are
Clebsch-Gordan coefficients.  $ F_\alpha(\alpha=x,y,z)$ are
$5\times5$ spin matrices obeying usual momentum commutation
relations $[F_\alpha,
F_\beta]=\varepsilon_{\alpha\beta\gamma}F_\gamma$.

For simplicity, we assume a uniform magnetic field applied along
the $z$-direction, and it is weak enough to ignore the quadratic
Zeeman effect. Expanding the field operators in the Wannier basis
and keeping only the lowest vibrational states,
$\Psi_\alpha=\sum_i b_{i\alpha} w({\bf r}-{\bf r}_i)$, Eq.
(\ref{h}) reduces to the generalized Bose-Hubbard Hamiltonian
\begin{equation}
H=-t\sum_{<i,j>}b_i^\dag b_j
-\mu\sum_i\hat{n}_i+\frac{c_0}{2}\sum_i\hat{n}_i(\hat{n}_i-1)
+\frac{c_1}{2}\sum_i(\hat{\bf {F}}_i^2-6\hat{n}_i)
+\frac{2c_2}{5}\sum_i\hat{S}_{i+}\hat{S}_{i-}-\sum_i p\hat F_{zi},
\label{b}
\end{equation}
where $ \hat {\bf F}_i=b_{i\alpha}^\dag {\bf
F}_{\alpha\beta}b_{i\beta}$, $\hat{n}_i=\sum_\alpha
b_{i\alpha}^\dag b_{i\alpha}$, $t=-\int d{\bf r}w_i^*({\bf
r})(-\hbar^2{\bf\nabla}^2/2m+V({\bf r}))w_j({\bf r})$ is the
hopping matrix element between adjacent sites $i$ and $j$,
$\mu=\bar{\mu}\int d{\bf r}|w_i({\bf r})|^2+\int d{\bf
r}w_i^*({\bf r})(-\hbar^2{\bf\nabla}^2/2m+V({\bf r}))w_i({\bf r})$
describes the effective chemical potential, and $c_i=\bar{c}_i\int
d{\bf r}|w_i({\bf r})|^4$ is on-site inter-atom interaction, where
the Hubbard approximation has been used to treat the multi-center
integral as a single-center one.
$\hat{S}_{i+}=\hat{S}_{i-}^\dag=(b_{i0}^\dag)^2/2-b_{i1}^\dag
b_{i-1}^\dag+b_{i2}^\dag b_{i-2}^\dag$ creates a spin-singlet
``pair''when applied to the vacuum, and we further assume
$p=g\mu_BB>0$ in the following discussion.

In the limit $c_0/t\longrightarrow \infty$, the hopping term can
be neglected, so the Hamiltonian is reduced to a diagonal matrix
with respect to sites. Then the single-site Hamiltonian is
\begin{equation}
h_0=-\mu\hat{n}+\frac{c_0}{2}\hat{n}(\hat{n}-1)+\frac{c_1}{2}(\hat{\bf
F}^2-6\hat{n})+\frac{2c_2}{5}\hat{S}_+\hat{S}_--p \hat F_z.
\label{c}
\end{equation}

For the sake of studying the quantum transition, the mean-field
approximation \cite{oosten} is used and the hopping term is
considered as a perturbation. Introducing the superfluid order
parameter $\phi_\alpha=<b_{i\alpha}>=\sqrt{n_{sf}}\zeta_\alpha$
($n_{sf}$ is the superfluid density and $\zeta_\alpha$ is a
normalized spinor $\zeta_\alpha^*\zeta_\alpha=1$), we decouple the
hopping term as $ b_{i\alpha}^\dag b_{j\alpha}\approx(\phi_\alpha
b_{i\alpha}^\dag+\phi_\alpha^*
b_{j\alpha})-\phi_\alpha^*\phi_\alpha$, then, the total hopping
term becomes the product of a site-independent term and the total
number of the sites. As a result, we only consider a single site
because the Hamiltonian of every site is identical in the
homogenous case, and the Hamiltonian is represented by a
site-independent effective Hamiltonian multiplied by the total
number of sites. The Hamiltonian of a single site reads
\begin{eqnarray}
&&h=h_0+h_1,\\
&&h_1=zt(\phi_\alpha b_\alpha^\dag+\phi_\alpha^*
b_\alpha-\phi_\alpha^*\phi_\alpha),\label{d}
\end{eqnarray}
where $h_1$ is the mean-field version of the hopping Hamiltonian
in the single site, and $z$ is the number of the nearest-neighbor
sites. When the ratio $c_0/t$ is very large, $h_1$ is considered
as a perturbation term.

\section{The Energy Eigenvalues and the Mott Ground States}
Before perturbative calculations, we solve the equation
\begin{equation}
h_0\psi=\varepsilon^{(0)}\psi
\end{equation}
to get the eigenvalues and eigenstates of $h_0$.
 In Eq. (\ref c), the operaters $\hat{S}_+$ and $\hat{S}_-$ satisfy
the $SU(1,1)$ commutation relations, namely $
[\hat{S}_z,\hat{S}_\pm]=\pm\hat{S}_\pm,\ \
[\hat{S}_+,\hat{S}_-]=-2\hat{S}_z$ together with
$\hat{S}_z\equiv(2\hat{n}+5)/4$,  and as a consequence, the
Casimir operator
$\hat{\textbf{S}}^2\equiv-\hat{S}_+\hat{S}_-+\hat{S}_z^2-\hat{S}_z$
commutes with $\hat{S}_\pm$ and $\hat{S}_z$ \cite{koashi,ueda}.
The eigenvalues of the mutual eigenstates for $\hat{\textbf{S}}^2$
and $\hat{S}_z$ are $\{S(S-1),S_z\}$ with $S=(2n_0+5)/4\ \
(n_0=0,1,2,...)$ and $S_z=S+n_s\ \ (n_s=0,1,2,...)$, which
guarantees that
$\hat{S}_+\hat{S}_-=\hat{S}_z^2-\hat{S}_z-\hat{S}^2$ is positive
semidefinite. The new quantum numbers $n_s$ and $n_0$ are
introduced and the operator $S_+$ raises $n_s$ by one while the
relation $n=2n_s+n_0$ holds, where $n$ is the total number of
bosons in a single site. The spin operator $\hat{\textbf{F}}$ and
the magnetic quantum number operator $\hat{F}_z$ commute with the
operators $\hat{S}_\pm$, and then, the eigenstates $\psi$ are
denoted as $|n_0,n_s,F,F_z;\lambda>$ where $\lambda$ labels
orthonormal degenerate states. The energy eigenvalue is given by
\begin{equation}
\varepsilon^{(0)}=-\mu
n+\frac{c_0}{2}n(n-1)+\frac{c_1}{2}[F(F+1)-6n]+\frac{2c_2}{5}n_s(n-n_s+\frac{3}{2})-pF_z,
\label{energy}
\end{equation}
So Mott states can be expressed as
$\prod_i|n_0,n_s,F,F_z;\lambda>_i$, where $i$ is lattice site
index. For the homogenous case, the zeroth order total energy is
$E^{(0)}=\sum_i \varepsilon^{(0)}=N_l\varepsilon^{(0)}$, where
$N_l$ is the number of lattice sites.

The energy eigenstates $|n_0,n_s,F,F_z;\lambda>$ can be
represented as \cite{ueda}
\begin{equation}
(\hat{F}_-)^{\Delta
F}(\hat{A}_0^{(2)\dag})^{n_{20}}\hat{P}_{(n_s=0)}(b_2^\dag)^{n_{12}}
(\hat{A}_2^{(2)\dag})^{n_{22}}(\hat{A}_0^{(3)\dag})^{n_{30}}(\hat{A}_3^{(3)\dag})^{n_{33}}|vac>,
\end{equation}
where
\begin{eqnarray}
\label{a1}
\hat{A}_0^{(2)\dag}&=&\frac{1}{\sqrt{10}}[(b_0^\dag)^2-2b_1^\dag
b_{-1}^\dag+2b_2^\dag b_{-2}^\dag],\\
\label{a2}
\hat{A}_2^{(2)\dag}&=&\frac{1}{\sqrt{14}}[2\sqrt{2}b_2^\dag
b_0^\dag-\sqrt{3}(b_1^\dag)^2],\\
\label{a3}
\hat{A}_0^{(3)\dag}&=&\frac{1}{\sqrt{210}}[\sqrt{2}(b_0^\dag)^3-3\sqrt{2}b_1^\dag
b_0^\dag
b_{-1}^\dag+3\sqrt{3}(b_1^\dag)^2b_{-2}^\dag+3\sqrt{3}b_2^\dag
(b_{-1})^2-6\sqrt{2}b_2^\dag b_0^\dag b_{-2}^\dag],\\
\label{a4} \hat{A}_3^{(3)\dag} &=&
\frac{1}{20}[(b_1^\dag)^3-\sqrt{6}b_2^\dag b_1^\dag
b_0^\dag+2(b_2^\dag)^2b_{-1}^\dag].
\end{eqnarray}
$P_{(n_s=0)}$ is the projection onto the subspace with $n_s=0$;
$n_{12},n_{20},n_{22},n_{30}=0,1,2,...,\infty$, $n_{33}=0,1$, and
$\Delta F=0,1,...,2F$ that are related to $\{n_0,n_s,F,F_z\}$ by
\begin{eqnarray}
n_0&=&n_{12}+2n_{22}+3n_{30}+3n_{33},\\
n_s&=&n_{20}, \\
F&=&2n_{12}+2n_{22}+3n_{33},\\
F_z&=&F-\Delta F.
\end{eqnarray}

From Eq. (\ref{energy}), we see that the minimum energy states
always satisfy $F_z=F$ when $p>0$. Thus, the problem of finding
the ground states reduces to minimizing the function:
\begin{eqnarray}
\varepsilon^{(0)}(F,n_s)&=&-\mu
n+\frac{c_0}{2}n(n-1)+\frac{c_1}{2}[(F-\frac{2p-c_1}{2c_1})^2-6n
-\frac{(c_1-2p)^2}{4c_1^3}]\nonumber\\
&&+\frac{2c_2}{5}n_s(n-n_s+\frac{3}{2}). \label{energypa}
\end{eqnarray}
The ground states of $h_0$ depend on the relation among
$c_1,c_2,p$ and $\mu$. When $n=1$, the ground state is
$|1,0,2,2;\lambda>$; when $n\geq 2$, there are four sorts of
ground states classified by different sign combinations of $c_1$
and $c_2$. Here the classification of the Mott ground states
differs from that of Ref. \cite{ueda}, which is labelled by
ferromagnetic, antiferromagnetic and cyclic phases. In general,
for $p>0$, when $c_1<0,\ c_2>0$ (ferromagnetic case), the third
and the fourth term in Eq. (\ref{energypa}) possess the minimal
values simultaneously, but when $c_1<0,\ c_2<0$ (ferromagnetic or
antiferromagnetic case), $c_1>0,\ c_2>0$ (ferromagnetic or cyclic
case) and $c_1>0,\ c_2>0$ (ferromagnetic or antiferromagnetic
case), the situation becomes complicated, since sometimes there
exists the competition between contributions of total spin and the
singlet ``pairs'' to eigenenergy. We list all Mott ground states
in detail in appendix.

\section{Phase Diagrams of Superfluid-Mott Insulator Transition}
From Section III, we know that $F=F_z$ always holds in the ground
states, while in the excited states this is not the case. As a
result of the applied magnetic field, the state degeneracy from
the different magnetic quantum numbers $F_z$ \cite{hou} is lifted.
More rich phase diagrams are expected than those in the absence of
the magnetic field.

To depict the phase diagrams, we consider the hopping term as the
perturbative one and calculate the first- and second-order
corrections to the ground energy, which are expressed as
\begin{equation}
\varepsilon_g^{(1)}=<g|h_1|g>=zt\sum_\alpha
\phi_\alpha^*\phi_\alpha,\ \ \ \alpha=-2,...,2,
\end{equation}
\begin{equation}
\varepsilon_g^{(2)}=\sum_{n\neq
g}\frac{|<g|h_1|m>|^2}{\varepsilon_g^{(0)}-\varepsilon_m^{(0)}}=\sum_{n\neq
g}\sum_\alpha\frac{z^2t^2|<g|b_\alpha+b_\alpha^\dag|m>|^2\phi_\alpha^*\phi_\alpha}
{\varepsilon_g^{(0)}-\varepsilon_m^{(0)}},\
\ \ \alpha=-2,...,2.
\end{equation}
Here $|g>$ denotes the ground state discussed in the appendix, and
$\{|m>\}$ represent excited states expressed as a cluster of
quantum numbers including $n_0,n_s,F,F_z,\lambda$. We can
calculate all the nonzero matrix elements of
$<g|b_\alpha+b_\alpha^\dag|m>$. Therefore, second-order
perturbation theory gives the form of the modified ground energy
as
\begin{equation}
\varepsilon_g=\varepsilon_g^{(0)}+\varepsilon_g^{(1)}+\varepsilon_g^{(2)}=\varepsilon_g^{(0)}+zt\sum_\alpha
A_\alpha(n,\tilde{\mu},\tilde{c}_0,\tilde{c}_1,\tilde{c}_2,\tilde{p})\phi_\alpha^*\phi_\alpha,\
\ \ \alpha=-2,...,2, \label{ge}
\end{equation}
where
$A_\alpha(n,\tilde{\mu},\tilde{c}_0,\tilde{c}_1,\tilde{c}_2,\tilde{p})$
is related to the first- and second-order corrections of the spin
component with magnetic quantum number $\alpha$ to the
zeroth-order ground energy. It depends on system parameters $n$,
$\tilde{\mu}$, $\tilde{c}_0$, $\tilde{c}_1$, $\tilde{c}_2$ and
$\tilde{p}$, where $\tilde{\mu}=\mu/zt,
\tilde{c}_0=c_0/zt,\tilde{c}_1=c_1/zt,\tilde{c}_2=c_2/zt,\tilde{p}=p/zt$
are dimensionless. Minimizing the ground energy function Eq.
(\ref{ge}), we find that $\phi_\alpha=0$ when
$A_\alpha(n,\tilde{\mu},\tilde{c}_0,\tilde{c}_1,\tilde{c}_2,\tilde{p})>0$
and $\phi_\alpha\neq 0$ when
$A_\alpha(n,\tilde{\mu},\tilde{c}_0,\tilde{c}_1,\tilde{c}_2,\tilde{p})<0$.
This means that
$A_\alpha(n,\tilde{\mu},\tilde{c}_0,\tilde{c}_1,\tilde{c}_2,\tilde{p})=0$
signifies the boundary between the superfluid and the Mott
insulator phases of the spin component with magnetic quantum
number $\alpha$.

\begin{figure}[ht]
 \includegraphics[width=0.8\columnwidth]{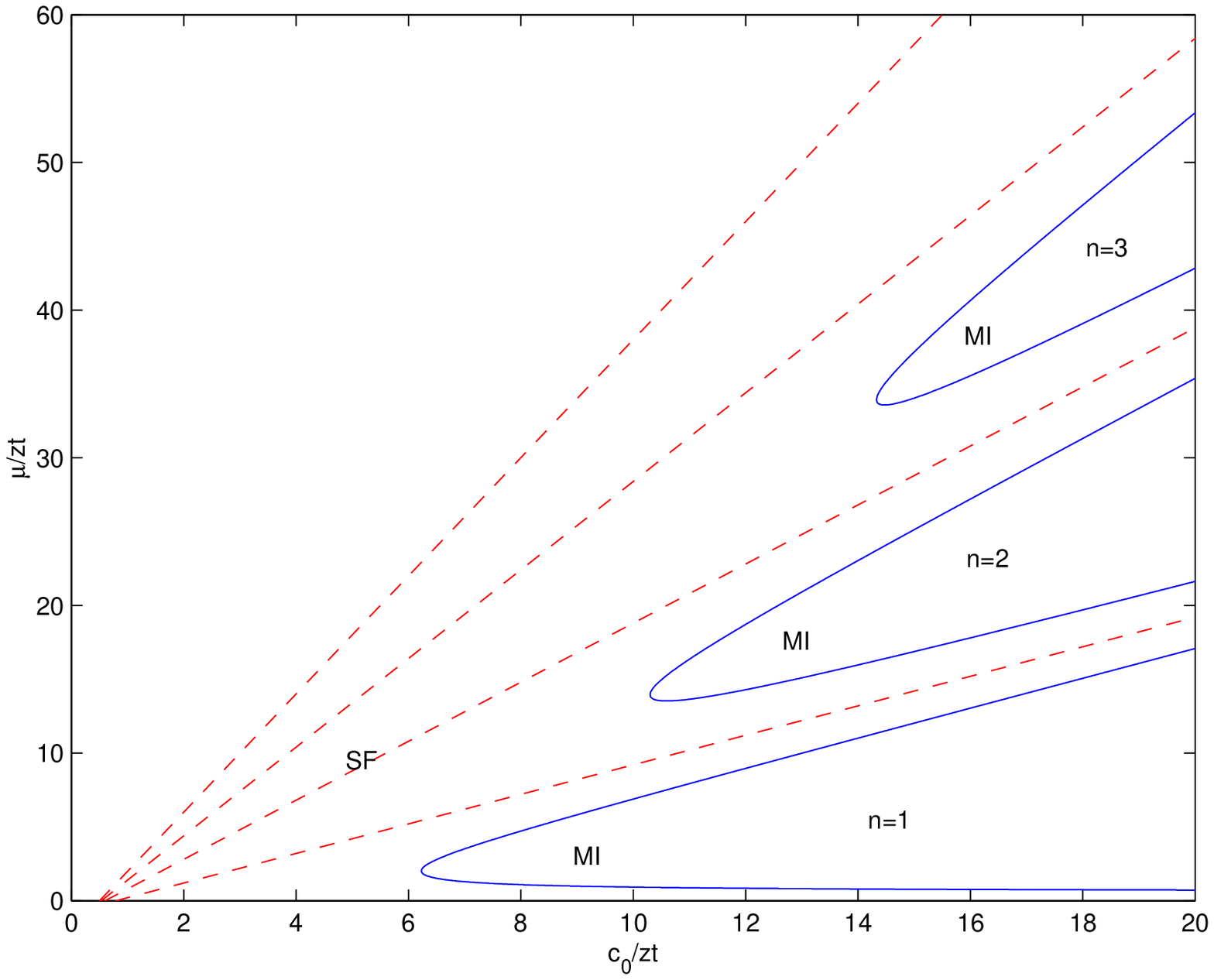}
 \caption{The phase diagram
of Bose-Hubbard Hamiltonian obtained from second-order
perturbation theory with solid lines for $c_1=-0.1zt$, $c_2=0.1zt$
and $p=0.2zt$. The dashed lines indicate the zeroth-order phase
diagram.}\label{1}
\end{figure}

\begin{figure}[ht]
\includegraphics[width=0.8\columnwidth]{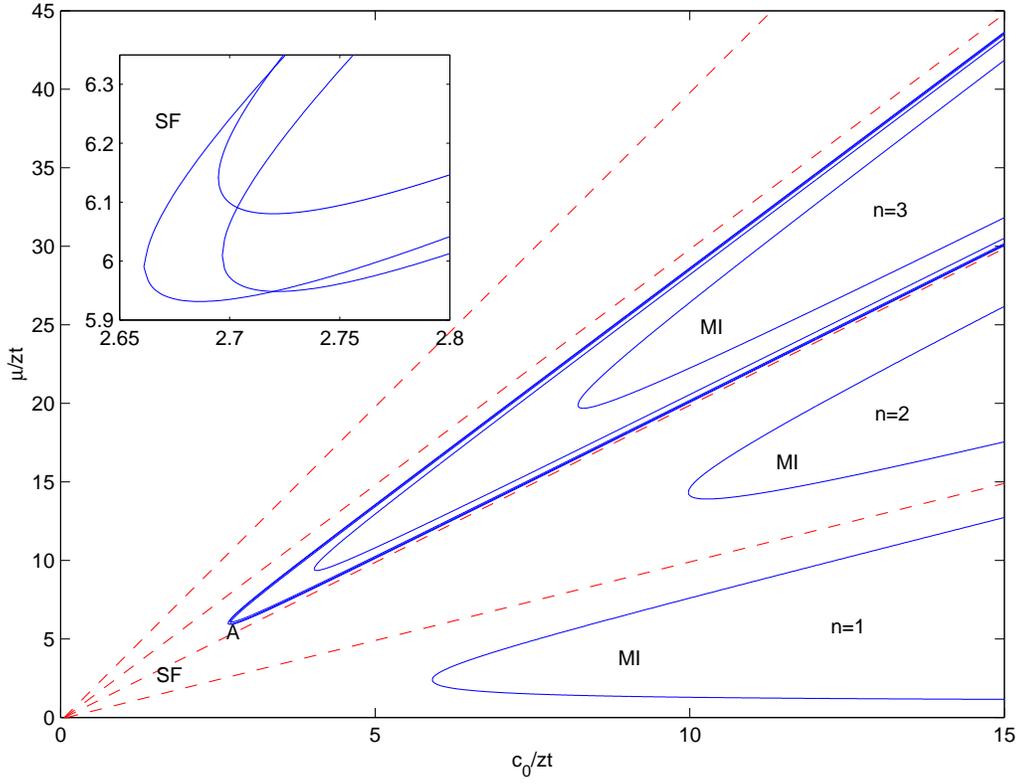} \caption{The same as in Fig.\ref{1}, but for $c_1=-0.02zt$,
$c_2=-0.25zt$ and $p=0.01zt$ with solid lines. For $n=3$, the
interior line is the phase boundary of spin component with Zeeman
level $m=2$; the middle line the phase boundary of spin component
with Zeeman level $m=-2$, and the external triple lines, which are
too close to be distinguished, the phase boundaries of spin
components with Zeeman levels $m=0,\pm1$. The inset shows an
expansion of the region labelled by A.} \label{2}
\end{figure}

\begin{figure}[ht]
\includegraphics[width=0.8\columnwidth]{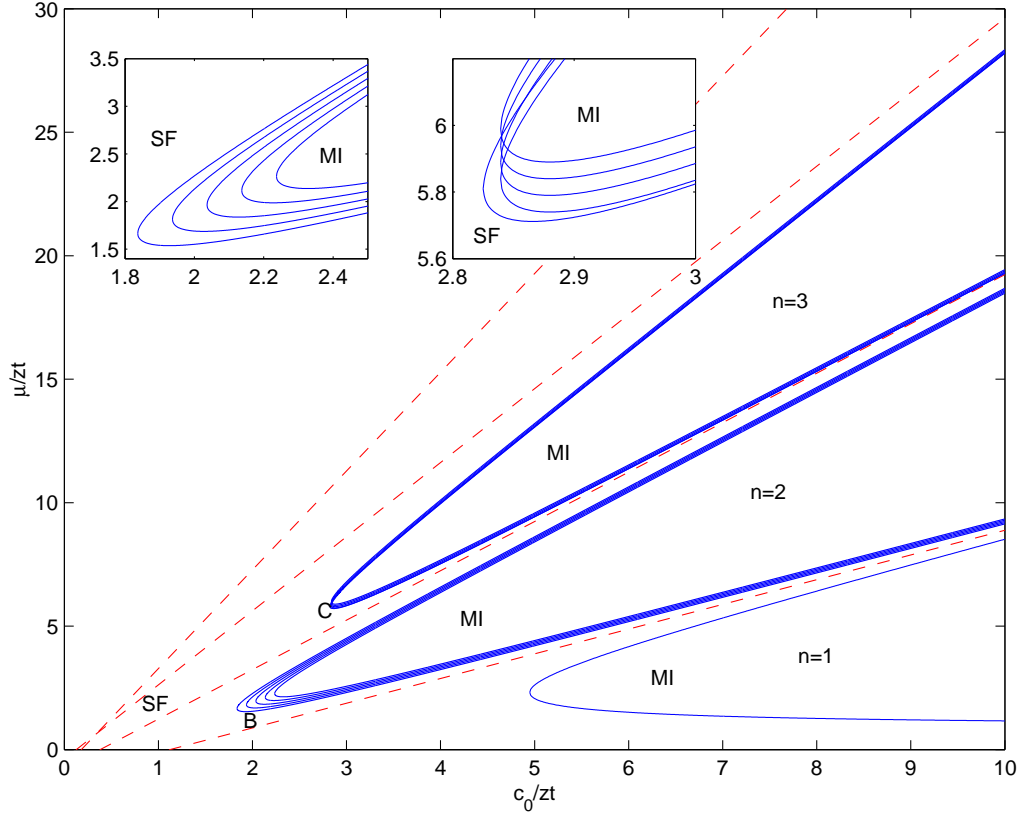}\caption{The same as in Fig.\ref{1}, but for $c_1=0.22zt$, $c_2=0.1zt$ and
$p=0.05zt$ with solid lines. When $n=2$, the region labelled by B
is enlarged in the left inset; the five lines represent the phase
boundaries of spin components with Zeeman levels $m=-2,-1,0,1,2$
respectively (from interior to external). For $n=3$, the five
lines are too close to be distinguished, so the region labelled by
C is enlarged in the right inset, in which the five lines the
phase boundaries of spin components with Zeeman levels
$m=0,\pm1,\pm2$. } \label{3}
\end{figure}

\begin{figure}[ht]
\includegraphics[width=0.8\columnwidth]{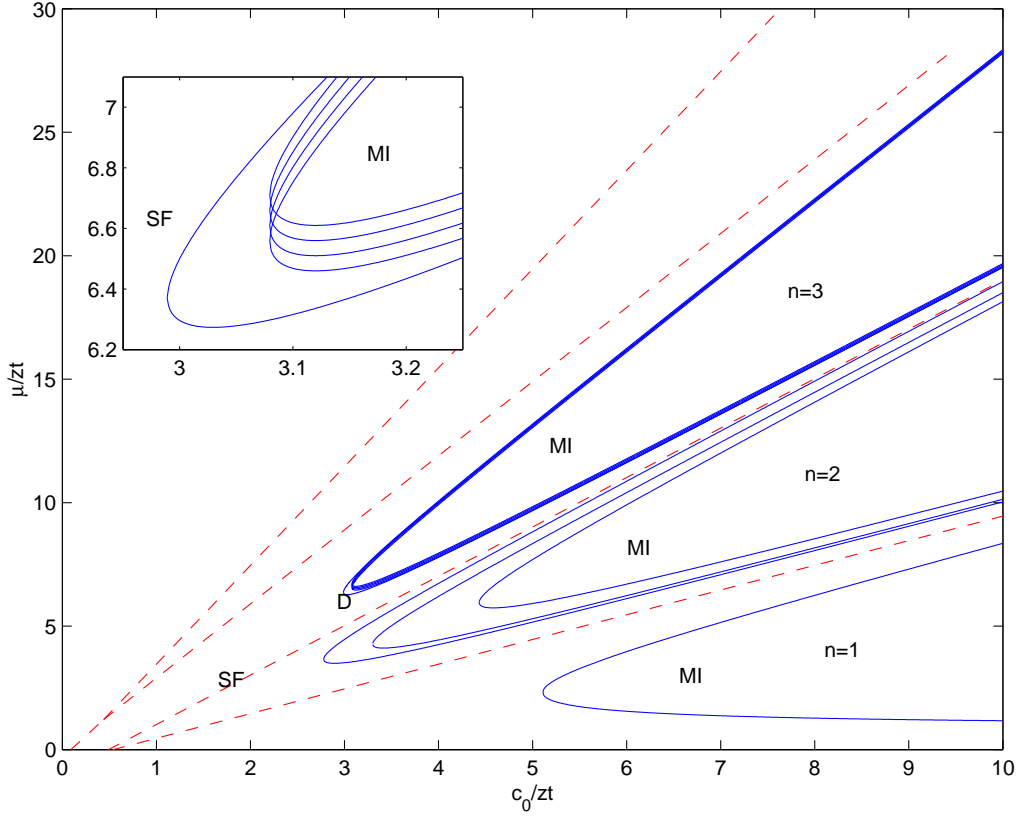} \caption{The same as in Fig.\ref{1}, but for $c_1=0.18zt$, $c_2=1.0zt$ and
$p=0.05zt$ with solid lines. For $n=2$, the three lines represent
the phase boundaries of spin components with Zeeman level
$m=1,0,2$ respectively (from interior to external). When $n=3$,
the five lines are too close to be distinguished, so the region
labelled by D is enlarged in the inset, in which the external line
is the phase boundary of spin component with Zeeman level $m=2$,
and the other four lines the phase boundaries of spin components
with Zeeman levels $m=0,\pm1,-2$.} \label{4}
\end{figure}

\begin{figure}[ht]
\includegraphics[width=0.8\columnwidth]{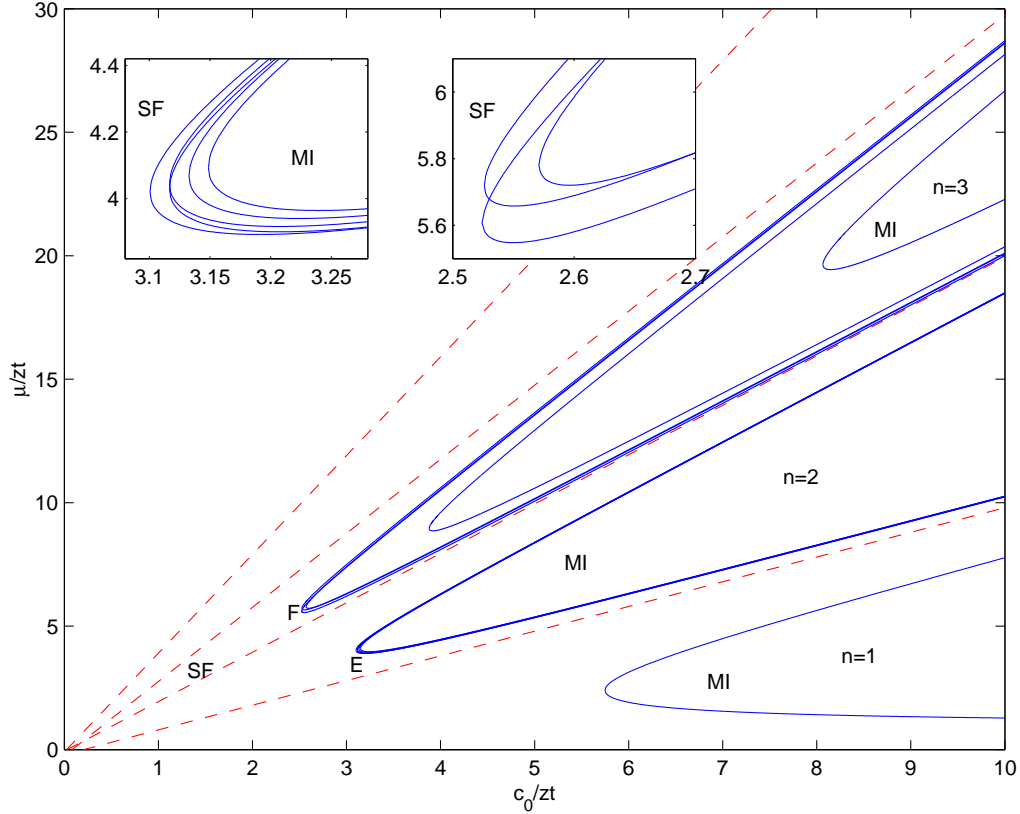}\caption{The same as
in Fig.\ref{1}, but for $c_1=0.02zt$, $c_2=-0.1zt$ and
$p=0.008zt$ with solid lines. When $n=2$, the five lines are too
close to be distinguished, so the region labelled by  E is
extended in the left inset, in which the interior line is the
phase boundary of spin component with Zeeman level $m=-1$; the
external line the phase boundary of spin component with Zeeman
level $m=1$; the three middle lines the phase boundaries of spin
components with Zeeman levels $m=0,\pm2$. For $n=3$, the exterior
line is the phase boundary of spin component with Zeeman level
$m=2$; the middle line the phase boundary of spin component with
Zeeman level $m=-2$, and the external triple lines, which are too
close to be distinguished, the phase boundaries of spin components
with Zeeman levels $m=0,\pm1$. The right inset shows an expansion
of the region labelled by F. } \label{5}
\end{figure}

Using the perturbation theory, we can analytically determined the
phase diagrams Fig. \ref{1}-\ref{5} for different cases. The phase
diagrams indicate that there exists a phase transition from Mott
insulator with integer number bosons to superfluid when the ratio
$c_0/t$ is decreased to a critical value. In the zeroth-order,
i.e., neglecting the hopping term, the ground state is Mott state
in which the occupation number per site is pinned at integer
$n=1,2,...$, corresponding to a commensurate filling of the
lattice. Different ground states may contain different spin
components. For example, there is only spin component with Zeeman
level $m=2$ when occupation number per site $n=1$; spin components
with $m=0,\pm 1,\pm 2$ for Mott state
$\prod_i(\hat{A}_0^{(2)}|0>)_i$, and spin components with
$m=0,1,2$ for $\prod_i(\hat{A}_2^{(2)}|0>)_i$. For the initial
Mott ground state including only one spin component, one
superfluid component occurs when lowering the ratio $c_0/t$, such
as the case $n=1,2,3$ in Fig. \ref{1}, $n=1,2$ in Fig. \ref{2},
and $n=1$ in Fig. \ref{3}-\ref{5}; for all the initial Mott ground
states containing multiple spin components, when lowering the
ratio $c_0/t$, multiple superfluid components appear, and the
phase boundaries between superfluid and Mott insulator phase for
different spin components are distinct, for instance, $n=3$ in
Fig. \ref{2}, $n=2,3$ in Fig. \ref{3}-\ref{5}. After analyzing the
phase diagrams, we find that the position of phase boundary is
related to average occupation number of spin component in the
initial Mott ground state, i.e., the larger the average occupation
number of spin component per site is, the easier the transition
from Mott insulator to superfluid phase. We also find that some
boundaries between superfluid and Mott insulator phases with
multi-spin components, such as $n=2$ in Fig. \ref{3},\ref{5} and
$n=3$ in Fig. \ref{3},\ref{4}, will turn to be identical when the
magnetic field vanishes. We can draw the conclusion that the
applied magnetic field results in some phase boundaries splitting.

\begin{figure}[ht]
\includegraphics[width=0.8\columnwidth]{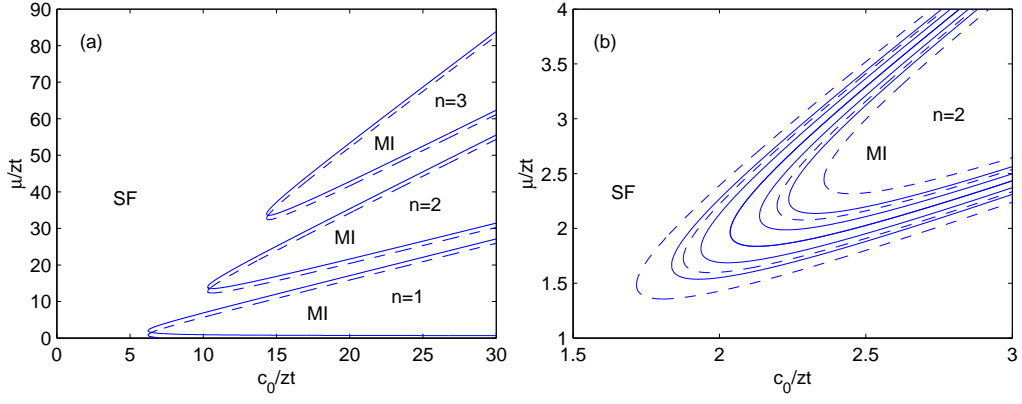}
\caption{The phase diagrams of Bose-Hubbard Hamiltonian obtained
from second-order perturbation theory for different $p$. In (a),
$c_1=-0.1zt$, $c_2=0.1zt$, with $p=0.2zt$ (solid lines) and
$p=0.8zt$ (dashed lines) respectively. In (b), when $n=2$,
$c_1=0.22zt$, $c_2=0.1zt$, with $p=0.05zt$ (solid lines) and
$p=0.08zt$ (dashed lines) respectively; the middle solid line
expresses the phase boundary of spin component with Zeeman level
$m=0$, and keeps invariant when $p=0.08zt$; the five solid lines
represent the phase boundaries of spin components with Zeeman
levels $m=-2,-1,0,1,2$ respectively (from interior to external);
the four dashed lines represent the phase boundaries of spin
components with Zeeman levels $m=-2,-1,1,2$ respectively (from
interior to external).} \label{6}
\end{figure}

Furthermore, the phase diagrams Fig. \ref{6}(a) and Fig.
\ref{6}(b) for different intensities of the applied field are
drawn. The position of some phase boundaries is related to the
intensity of the applied magnetic field. In Fig. \ref{6}(a), when
the applied magnetic field increases, for the same Mott ground
state containing only one spin component with Zeeman level $m=2$,
the phase diagrams will shift along the direction with chemical
potential decreasing. In Fig. \ref{6}(b), for the same Mott ground
state containing spin components corresponding to Zeeman levels
$m=0,\pm1,\pm2$, when the magnetic field increases, the phase
boundaries of spin components with Zeeman levels $m=\pm1,\pm2$
move, but the phase boundary of spin component with Zeeman level
$m=0$ keeps invariant. Moreover, one can see that for positive and
negative Zeeman levels, the phase boundaries between SF and MI
will move in opposite directions, and for the spin component with
positive Zeeman level, the transition from MI to SF becomes easier
when the applied magnetic field increases.

\section{Remarks and Conclusion}
For simplicity, in this paper we assume the uniform magnetic field
is applied along the $z$-direction, and $p$ is positive. In fact,
the analysis of the Mott ground states for $p>0$ is enough since
the sign of $p$ does not alter the physics. When $p<0$, the Mott
ground states satisfy $F_z=-F$, and have the same form as those in
the case of $p>0$. We find that the influence of magnetic field on
the phase diagrams is manifold, and the boundaries between SF and
MI are essentially dependent on the magnetic properties of the
ground states. This work only discuss the phase diagrams
corresponding to different magnetic fields with the same Mott
ground state, i.e., the continuous change of the phase boundaries.
The case that Mott ground states change when the magnetic field
increases, i.e., the sudden jump of the phase boundaries, is
beyond the scope of the present paper and is not shown. For
instance, when $c_1>0,\ c_2>0$, if $p$ increases from $0$ to the
value large enough to satisfy $p/(2n+1)>c_1$, the magnetization
$F$ can jump from the minimum to the maximum one. They are the
subject of future study. In addition, it is worth to note that
phase diagrams of the zero magnetic field cannot be derived by
taking $p=0$ simply, since our derivation is based on the
degeneracy lifting.

In conclusion, we have investigated the quantum phase transition
from Mott insulator to superfluid phase of spin-2 cold bosons with
repulsive interaction in optical lattices under the influence of a
uniform magnetic field at zero temperature. The phase diagrams
show that the system undergoes a phase transition from Mott
insulator with integer number bosons at each site to superfluid
phase when the ratio $c_0/t$ is decreased to a critical value.
Different Mott ground states may contain different spin
components. The position of phase boundary is related to average
occupation number of spin component in the initial Mott ground
state. For the initial Mott ground state including only one spin
component, one superfluid component appears when lowering the
ratio $c_0/t$. For some Mott ground states with multiple spin
components, the applied magnetic field leads to the splitting of
the phase boundaries, so that the phase boundaries between
superfluid and Mott-insulator phase for different spin components
are distinct in all ground states. In particular, we draw the
phase diagrams corresponding to different intensities magnetic
field for the initial Mott ground state containing one-spin and
multi-spin components. They qualitatively show the way of the
phase boundaries' moving with the intensity of the applied
magnetic field. It is found that the phase boundaries can be moved
by only changing the intensity of the applied magnetic field. For
the spin component with positive Zeeman level, the larger
intensity of the magnetic field is, the easier the transition from
MI to SF happens. These theoretical results are expected to be
practically helpful to the experimental study of the field-tuned
SF-MI transition of bose atoms with hyperfine spin in an optical
lattice.

\begin{acknowledgments} This work is in part supported by
NSF of China Grant No.A0124015.
\end{acknowledgments}

\appendix*
\section{List of the Mott Ground States}
\begin{enumerate}
%1
\item{$c_1<0,\ \ c_2>0$. Because $(2p-c_1)/(2c_1)\leq0$, the third
term and the fourth term in Eq. (\ref{energypa}) have the minimal
values at the same time when $F=2n$ and $n_s=0$. Hence, the ground
state is $|n,0,2n,2n;\lambda>$.}

%2
\item{$c_1<0,\ \ c_2<0$. The relation $(2p-c_1)/(2c_1)\leq0$
stands, and the competition between contributions of total spin
and the singlet ``pair'' happens in the premise of
$\frac{F}{2}+2n_s=n$. If we skip over the fact for the moment that
the singlet ``pair'' number $n_s$ is an integer, then the
condition to minimize the energy function is given by}
\begin{equation}
n_s=\frac{10c_1(4n+1)-c_2(2n+3)-20c_1p}{80c_1-4c_2}.
\end{equation}
Because the singlet ``pair'' number must be an integer, we write
$n_s$ in terms of the closest integer number $n_s^0$ and the
decimal part, i.e. $n_s=n_s^0+\alpha$, where the number $\alpha$
satisfies $-1/2<\alpha<1/2$, which can be rewritten as,
  \begin{equation}
  \label{n}
   n_s^0-\frac{1}{2}<\frac{10c_1(4n+1)-c_2(2n+3)-20c_1p}{80c_1-4c_2}<n_s^0+\frac{1}{2}.
   \end{equation}
\begin{enumerate}
  \item{Because the singlet ``pair'' number is not negative, $n_s$ must be zero for $n_s^0\leq0$. Hence,
  the ground state is $|n,0,2n,2n;\lambda>$.}
  \item{For $0<n_s^0<\frac{n}{2}$, the eigenenergy is lower when the singlet ``pair'' number is $n_s^0$
  than any other integer. So the ground state is $|n-2n_s^0,n_s^0,2n-4n_s^0,2n-4n_s^0;\lambda>$.}
  \item{For $n_s^0\geq\frac{n}{2}$, the singlet ``pair'' number takes the highest value as it can.
  (i)When $n$ is even, the ground state is $|0,n/2,0,0;\lambda>$;
   (ii)when $n$ is odd, the ground state is $|1,(n-1)/2,2,2;\lambda>$.}
   \end{enumerate}

%3
\item{$c_1>0,\ \ c_2>0$. For convenience, we introduce a new
parameter
\begin{eqnarray}
\label{f0} F_1=\frac{2p-c_1}{2c_1}=F_0+\alpha,
\end{eqnarray}
where $F_0$ is the integer number which is the closest integer to
$F_1$.} In Eq. (\ref{energypa}), when $n_s=0$ the fourth term has
the minimal value, and if total spin $F$ takes the appropriate
integer the third term possesses the minimal one.
\begin{enumerate}
\item{When $F_0\leq0$, i.e., $c_1>p$, the energy eigenvalue has
its minimum if $F=0$ and $n_s=0$. However, we must notice some
special cases because there have some forbidden values
\cite{koashi,ueda}, that is, $F=1,2,5,2n_0-1$ are not allowed when
$n_0=3k(k\in Z)$, and $F=0,1,3,2n_0-1$ are forbidden when
$n_0=3k\pm 1(k\in Z)$.}
  \begin{enumerate}
  \item{For $n=3k(k\in Z)$, $|n,0,0,0;\lambda>$ with $n_s=0$
  and $F=0$
  is the ground state.}
  \item{For $n=3k-1(k\in Z)$, the state with $n_s=0$ and $F=0$
  is not allowed simultaneously. If $n_s=0$, the lowest allowed
  value of total spin $F$ is $2$. On the other hand, $F=0$ is not forbidden
  when $n_0$ is $3k(k\in Z)$. $n_s=1$ is the lowest value satisfying the condition due to $n=2n_s+n_0$. So
  the state with $n_s=1$ and $F=0$ is a possible ground state.
  Comparing both eigenenergies for the two cases $F=0,n_s=1$ and
  $F=2,n_s=0$, we get the ground state, (i)$|n-2,1,0,0;\lambda>$
  for $c_2<(15c_1-10p)/(2n+1)$ and (ii) $|n,0,2,2;\lambda>$ for
  $c_2>(15c_1-10p)/(2n+1)$.}
   \item{For $n=3k+1(k\in Z)$, the
   competition exists between contributions of total spin and the singlet
   ``pair''
   to eigenenergy. But if $F=0$, $n_s$ is at least $2$. Therefore, the ground state is
   (i)$|n-4,2,0,0;\lambda>$ for $c_2<(15c_1-10p)/2(2n-1)$; (ii)
   $|n,0,2,2;\lambda>$ for $c_2>(15c_1-10p)/2(2n-1)$.}
 \end{enumerate}
 \item{When $F_0\geq2n$, i.e., $c_1<p/(2n+1)$, the ground
 state is
  $|n,0,2n,2n;\lambda>$ with $F=2n$ and $n_s=0$.}
 \item{When $0<F_0<2n$, i.e., $p/(2n+1)<c_1<p$, the eigenenergy
 is thus lower when $F$ is closer to $F_1$ and when
$n_s$ is smaller. Except for the case of $F_0$ taking the
forbidden values of $F$, the ground state is
$|n,0,F_0,F_0;\lambda>$.} When $F_0$ equals to the forbidden
values of $F$, $F$ may take the allowed integer next-nearest
  to $F_1$, i.e.,
  $F_0\pm1$ or $F_0\pm2$, or at the cost of increasing $n_s$ to $1$ or $2$,
  since any of the three values $0,1,2$ of $n_0$ mod $3$ is realized by setting
  $n_s$ as 0,1,or 2 owing to the relation $n=2n_s+n_0$.
  Whether or not the states $|n,0,F_0\pm1,F_0\pm1;\lambda>$, $|n,0,F_0\pm2,F_0\pm2;\lambda>$, $|n-2,1,F_0,F_0;\lambda>$,
  and $|n-4,2,F_0,F_0;\lambda>$ can be the lowest-energy state depends on the ratio $c_2/c_1$.
\end{enumerate}

%4
\item{$c_1>0,\ \ c_2<0$. The eigenenergy is the
lowest if $F=F_0$ and $n_s$ at its highest value. However, these
two choices are not always satisfied simultaneously.
\begin{enumerate}
\item{$F_0\leq0$, i.e., $c_1>p$. }
  \begin{enumerate}
  \item{When $n$ is even, $|0,n/2,0,0;\lambda>$is the ground state.}
  \item{When $n$ is odd, $n_s$ has the highest value $(n-1)/2$. But $F$ is not zero  when $n_s=(n-1)/2$.
  Alternatively, there is another case that $F=0$ and $n_s=(n-3)/2$. Hence, the ground state is (i) $|1,(n-1)/2,2,2;\lambda>$
   for $c_1<(7|c_2|+10p)/15$; (ii) $|3,(n-3)/2,0,0;\lambda>$ for $c_1>(7|c_2|+10p)/15$.}
\end{enumerate}
\item{$F_0>0$, i.e., $c_1<p$.} Note that
$n_s^0$ is shown in Eq. (\ref{n}), similar to the analysis of case
2., it is divided into three cases.
\begin{enumerate}
\item{For $n_s^0\leq0$, the ground state is
$|n,0,2n,2n;\lambda>$.}
  \item{For $0<n_s^0<\frac{n}{2}$, the ground state is $|n-2n_s^0,n_s^0,2n-4n_s^0,2n-4n_s^0;\lambda>$.}
  \item{For $n_s^0\geq\frac{n}{2}$,
  (i)when $n$ is even, $|0,n/2,0,0;\lambda>$ is the ground state;
   (ii)when $n$ is odd, $|1,(n-1)/2,2,2;\lambda>$ does.}
\end{enumerate}

  \end{enumerate}}

\end{enumerate}

\newpage
%f1

%f2

%f3

%f4

%f5

%f6
\end{document}